\documentclass[prb,superscriptaddress,onecolumn]{revtex4}

\usepackage{physics}
\usepackage{graphicx}
\usepackage{dcolumn}
\usepackage{bm}
\usepackage{siunitx}
\usepackage{booktabs}
\usepackage{amssymb}
\usepackage{amsfonts}
\usepackage[english]{babel}
\usepackage{hyperref}
\usepackage{bigints}
\usepackage{eqnarray,amsmath}
\usepackage{textcomp}

\usepackage{amsmath,amssymb}
\usepackage{lipsum}

\usepackage[usenames, dvipsnames]{color}

\usepackage{epstopdf}

\begin{document}

\title
  {Tuning the dynamics of chiral domain walls of ferrimagnetic films with the magneto-ionic effect }

\author{Cristina Balan}
\author{Jose Pe$\tilde{\mathrm{n}}$a Garcia}
\author{Aymen Fassatoui}
\author{Jan Vogel}
\author{Dayane de Souza Chaves}
\affiliation{Univ. Grenoble Alpes, CNRS, Institut N\'eel, 38042 Grenoble, France}
\author{Marlio Bonfim}
\affiliation{Dep. de Engenharia Elétrica,
Universidade Federal do Parana, Curitiba, Brasil}
\author{Jean-Pascal Rueff}
\affiliation{Synchrotron SOLEIL, L'Orme des Merisiers, Saint-Aubin, 91192 Gif-sur-Yvette, France}
\author{Laurent Ranno}
\author{Stefania Pizzini}
\affiliation{Univ. Grenoble Alpes, CNRS, Institut N\'eel, 38042 Grenoble, France}
\email{stefania.pizzini@neel.cnrs.fr}

\begin{abstract}
The manipulation of magnetism with a gate voltage is expected to lead the way towards the realization of energy-efficient spintronics devices and high-performance magnetic memories.  Exploiting magneto-ionic effects under micro-patterned electrodes in solid-state devices adds the possibility to modify magnetic properties locally, in a non-volatile and reversible way. Tuning magnetic anisotropy, magnetization and
Dzyaloshinskii-Moriya interaction allows modifying ``at will'' the dynamics of non trivial magnetic textures such as skyrmions and chiral domain walls in magnetic race tracks.
In this work, we illustrate efficient magneto-ionic effects in a ferrimagnetic Pt/Co/Tb stack using a  ZrO$_2$ thin layer as a solid state ionic conductor. When a thin layer of terbium is deposited on top of cobalt, it acquires a magnetic moment that aligns antiparallel to that of cobalt, reducing the effective magnetization.  Below the micro-patterned electrodes, the voltage-driven migration of oxygen ions in a ZrO$_2$ towards the ferrimagnetic stack partially oxidizes the  Tb layer, leading  to the local variation not only of the spontaneous magnetization, but also of the effective magnetic anisotropy and  of the Dzyaloshinskii-Moriya interaction. This leads to a huge increase of the domain wall velocity, which varies from 10 m/s in the pristine state to 250 m/s after gating. This non-volatile and reversible tuning of the domain wall dynamics may lead to applications to reprogrammable magnetic memories or other spintronic devices.
\end{abstract}

\maketitle

The study of domain wall (DW)  dynamics in magnetic thin films is receiving a wide interest not only for its rich physics but also in view of applications to spintronic devices. The discovery that extremely large velocities can be reached by chiral N\'{e}el DWs stabilized by the interfacial Dzyaloshinskii-Moriya interaction (DMI) \cite{Thiaville2012} driven either by magnetic fields or by spin-orbit torques \cite{Miron2010,Miron2011,Ryu2013,Emori2013,Pham2016} has widened the interest for this field. Recent studies have shown that electric fields provide an efficient route to tune the DW dynamics through the modification of interfacial magnetic properties, and in particular interfacial magnetic anisotropy. Relatively small variations of the magnetic anisotropy obtained when the gate voltage is at the origin of electronic effects (charge accumulation/depletion) can induce large variations of the DW velocities in the thermally activated (or creep) regime, where the DW velocity depends exponentially on the DW energy \cite{Schellekens2012,Bauer2012a,Bernand-Mantel2013,Chiba2012,Franke2015}.  Remarkably, Koyama et al. \cite{Koyama2018} observed changes of DW velocity exceeding 50 m/s in the depinning regime in Pt/Co/Pd/MgO stacks, using an HfO$_2$ dielectric layer.   The electric field effect on the magnetic anisotropy, which is volatile when the voltage induces only electronic charge modifications, can on the other hand be persistent when the chemical nature of the material is tuned via the migration of ionic species within solid state ionic conductors (GdOx, HfO$_2$, ZrO$_2$, ..)  \cite{Bauer2013,Bi2014,Bauer2015,Zhou2016,Fassatoui2020,Fassatoui2021}. In this case, much larger variations of the magnetic anisotropy have been observed, allowing to switch the easy magnetization axis from out-of-plane to in-plane with gating. So far, rare are the works showing important electric field driven variations of DW velocities beyond the thermally activated regime, where the DW velocities can largely exceed  100 m/s in systems with DMI \cite{Lin2016}. Since in this regime the DW velocities depend linearly on the magnetic parameters (magnetization, anisotropy or DMI) much smaller variations are expected. So far, large variations of current-driven DW velocities (up to $\pm$ \SI{100}{\meter/\second}) have been observed only  for a synthetic antiferromagnetic stacks, where the remanent magnetization was tuned by a controlled oxidation of the upper magnetic layer driven by ion migration \cite{Guan2021}. In that work a ionic liquid was used as ion conductor leading to a non-local modification of the magnetic properties and typical gating times of several tens of minutes.

In this work we have used the magneto-ionic effect to tune \textit{locally} the field-driven DW velocity in a ferrimagnetic Pt/Co/Tb/Al stack, using a ZrO$_2$ dielectric film as oxygen ion conductor. We demonstrate that the large variation of the DW velocity  results from the tuning  of not only  the magnetic anisotropy but also of the effective magnetization and DMI constant which result from the partial oxidation of the Tb layer. Using ultrashort magnetic field pulses, the domain wall dynamics could be  studied for magnetic fields up to $B_{\mathrm{z}}$=\SI{400}{\milli\tesla}.  DW velocities driven by the same magnetic  field strength, were observed to change from a few \si{\meter/\second} in the pristine state, to more than \SI{200}{\meter/\second} after the application of the gate voltage.

\begin{figure}[htbp]
    \centering
    \includegraphics[scale=0.7]{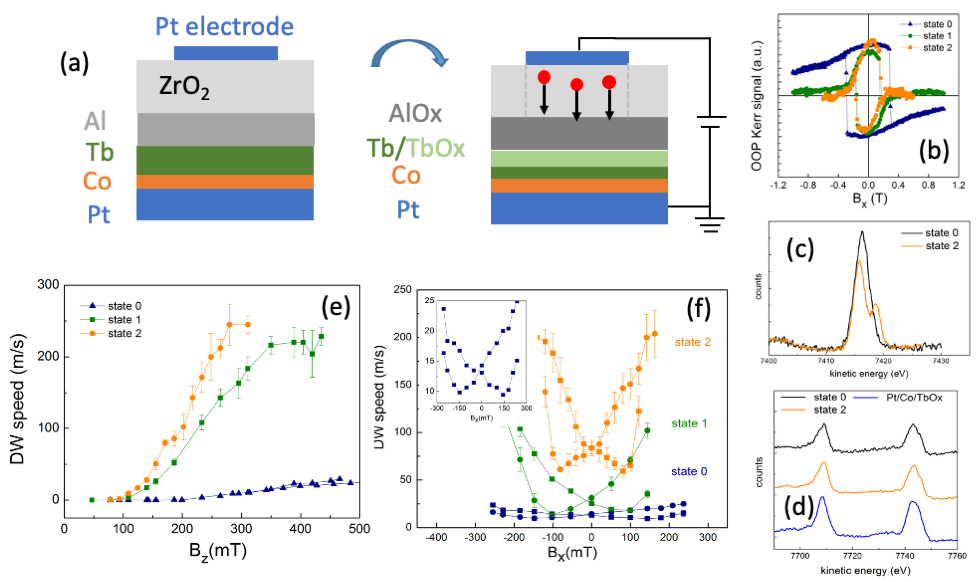}
    \caption{\textbf{Effect of a negative gate voltage on magnetic parameters and DW dynamics}; (a): sketch of the multilayer stack and effect of the migration of oxygen ions on the Al and Tb layers; (b) polar Kerr hysteresis loops vs. in-plane magnetic field $B_x$  before and after application of a negative gate voltage;
    (c-d) HAXPES spectra at the Al 1s edge (c) and Tb 4s edge (d)   before and after the application of the gate voltage; (e) DW velocity vs. $B_z$ field  modified by the gate voltage ; (f): DW velocity vs. in-plane field $B_x$  showing the modification of the $H_{DMI}$ field after gating.   }
    \label{fig:1}
\end{figure}

\subsection{Magneto-ionic effect on magnetic parameters and DW dynamics}

Ferrimagnetic Pt(4)/Co(1)/Tb(0.8)/Al(3) magnetic stacks, (Fig.  \ref{fig:1}(a)) were deposited by magnetron sputtering on Si/SiO$_2$ wafers  and patterned into micrometric strips. A ZrO$_2$ dielectric layer was deposited on top of the whole sample. The capacitor-like structures were then completed by the addition of local Pt gates (Fig.  \ref{fig:1}(b))
(see Methods and also \cite{Fassatoui2020,Fassatoui2021}).

The stacks display a perpendicular magnetization  as shown via \textit{M-H} hysteresis loops measured under an external out-of-plane magnetic field. The small value of the spontaneous magnetization ($M_\mathrm{s}=$  \SI{0.4}{\mega\ampere/\meter}) and the presence of a compensation temperature at around \SI{100}{\kelvin} confirm the ferrimagnetic nature of the samples, in which the magnetic moments of Tb align antiparallel to those of  Co.

The magnetic properties are locally modified below the Pt electrodes \textit{via} the application of a gate voltage, $V_{\mathrm{g}}$. The field-driven DW motion was studied in the region below the Pt electrodes using polar magneto-optical effect  Kerr
effect (MOKE) microscopy, both in the pristine state and after the application of the gate voltage. In the pristine state, the DW dynamics is the same below and outside the electrode.

In the pristine state (hereafter ``\textit{state 0}''), a large anisotropy field ($B_{\mathrm{k}}$=\SI{1.7}{\tesla})  is measured with MOKE, similar to the one measured by  VSM-SQUID before patterning (Fig.\ref{fig:1}(b)). In ``\textit{state 0}'' the DW velocity shows a very slow increase with the out-of-plane magnetic field, reaching only \SI{10}{\meter/\second} for $B_{\mathrm{z}}$=\SI{300}{\milli\tesla} (Fig.\ref{fig:1}(e)). The anisotropic motion of the DW, driven by a field $B_{\mathrm{z}}$ in the presence of a static in-plane magnetic field normal to the DW plane, confirms the presence of anti-clockwise chiral N\'{e}el DWs and allows us to determine the value of the DMI field \cite{Je2013}. Such field is related to the DMI constant $D$ through the relation $\mu_0 H_{DMI}=D/(M_{s}\Delta)$, where $\Delta=\sqrt{A/K_{eff}}$ is the DW parameter, $A$ is the exchange stiffness and  $K_{eff}=K_u - 1/2 \mu_0 M_s^2$ the effective anisotropy. In the  pristine state $\mu_0H_{DMI}$= \SI{150}{\milli\tesla} (Fig. \ref{fig:1}(f)).

After gating for the first time with a negative voltage (see Methods), in the so-called ``\textit{state 1}'', we observe a strong decrease of both the anisotropy field ($B_K$=\SI{300}{\milli\tesla}) and the DMI field  ($\mu_0H_{\mathrm{DMI}}$=\SI{100}{\milli\tesla}). This is accompanied by a huge increase of the DW velocity in the large magnetic field regime ($v$=\SI{140}{\meter/\second} for $B_{\mathrm{z}}=$\SI{300}{\milli\tesla}).

Under the application of a second negative gate voltage, in the so-called ``\textit{state 2}'' a further decrease of the anisotropy field and of the DMI field is also observed (see Table I).  The
DW velocity increases and reaches \SI{250}{\meter/\second}, corresponding to an unprecedented 2400\% variation of the DW speed in the flow regime.

In addition to the strong variation of the maximum domain wall velocity, a significant decrease of the depinning field, $H{_{dep}}$, i.e.,  the field for which the DW overcomes the energy barrier associated to the disorder, is also observed after gating. From Ref. \cite{Jeudy2018}, $H_{dep} \propto ~\sigma/M_{\mathrm{s}}$, where $\sigma=4\sqrt{AK_{eff}}-\pi D $ is the DW energy density. Therefore, the observed decrease of the depinning field after gating is consistent with the observed decrease of the anisotropy field and the increase of the spontaneous magnetization that will be discussed in the following sections.

\begin{figure}[htt]
    \centering
    \includegraphics[scale=0.65]{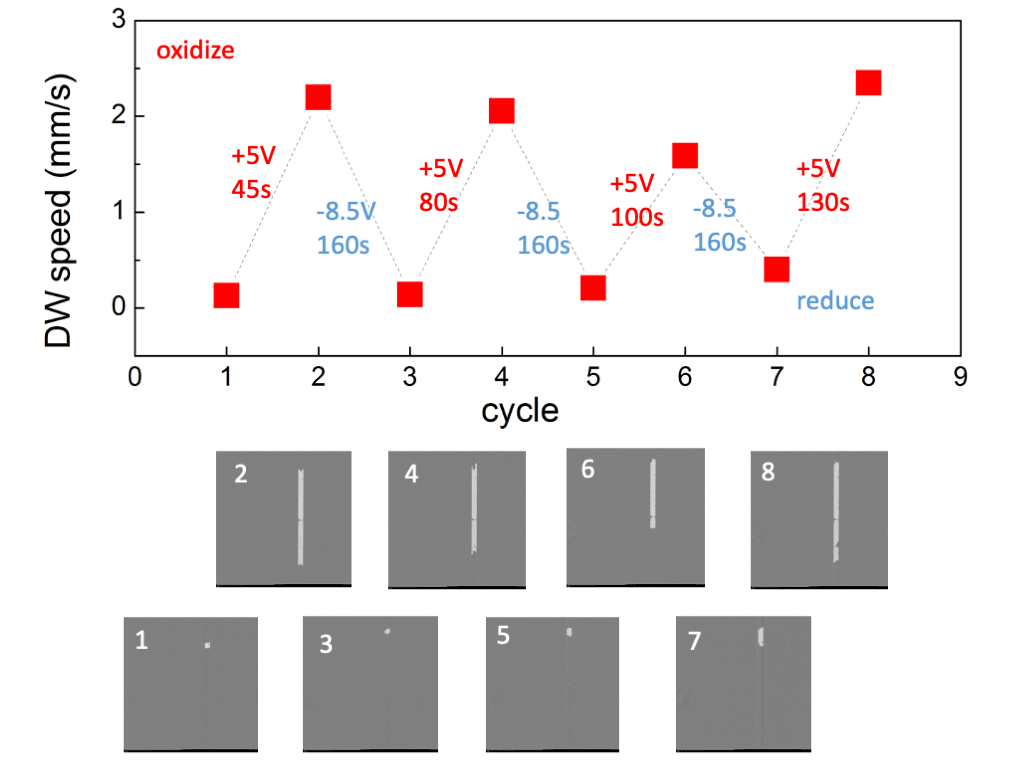}
    \caption{\textbf{Reproducible tuning of the DW velocity by electric field }. The domain walls are driven by a magnetic field $B_z$=15mT using a standard electromagnet (see Methods). The white contrast in the differential Kerr microscopy images represents the displacement of the DW below the Pt electrode, driven by the application of a 20ms long magnetic field pulse. }
    \label{fig:reproducibility}
\end{figure}

The modification of the DW velocity after gating is non volatile, as the magnetic state stabilized by electric field persists for several weeks after the removal of the gate voltage. The effect is also reversible as can be seen in Fig. \ref{fig:reproducibility}, where we report the DW velocity driven by a magnetic field $B_z$=15mT, measured after the application of successive positive and negative voltages. Two states with low and high velocity are clearly observed, the ratio between the two being at least of a factor 15 for this applied magnetic field.

\subsection{Effect of magnetic parameters on the field-driven DW motion}

\begin{figure}[b!]
    \centering
   \includegraphics[scale=0.65]{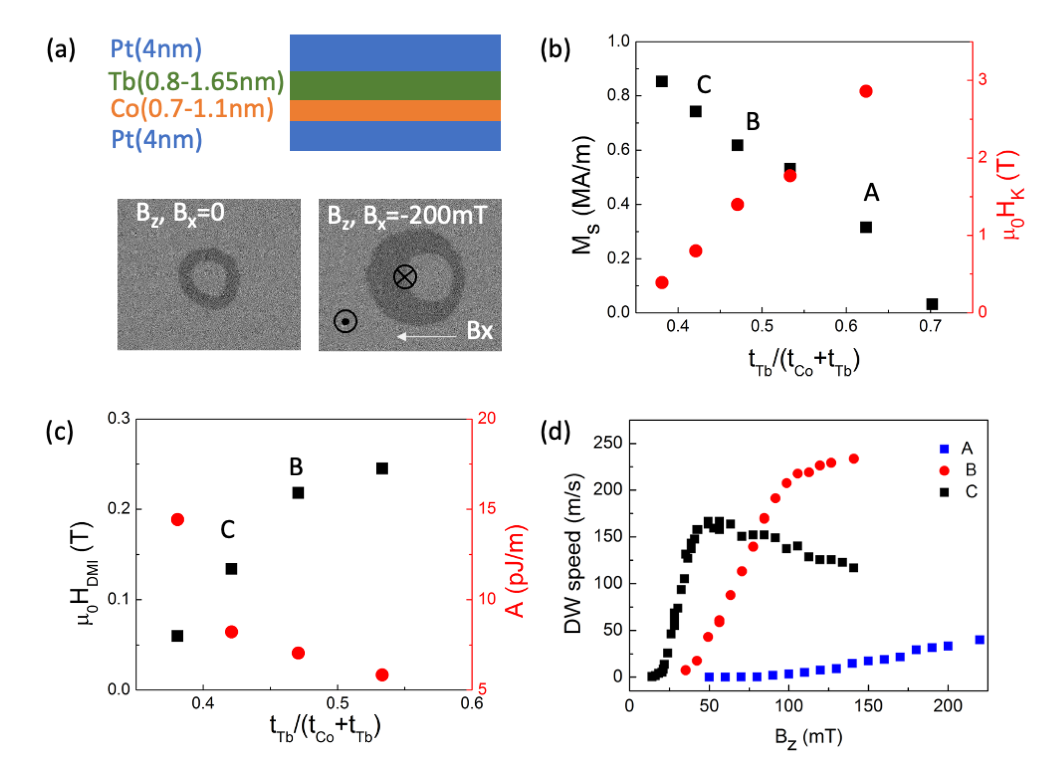}
    \caption{\textbf{Magnetic parameters and domain wall velocities  for the Pt/Co/Tb reference samples}; (a): sketch of the reference samples stacks and differential Kerr microscopy images showing the isotropic motion of the DWs driven by a field $B_z$ and the anisotropic motion in the presence of a static in-plane field, reflecting the presence of DMI;  (b) spontaneous magnetization and in-plane anisotropy field vs. $t_{eff}=\frac{t_{Tb}}{t_{Co}+t_{Tb}}$; (c) $H_{DMI}$ field and exchange stiffness as a function of
    $t_{eff}$; (d)      DW velocity vs. $B_z $ in Pt/Co/Tb reference samples A, B and C.}
    \label{fig:2}
\end{figure}

In our previous works \cite{Fassatoui2020,Fassatoui2021} we have shown that in capacitor-like devices prepared in the same way as the one studied here, the ZrO$_2$ dielectric layer deposited on top of  Pt/Co/MOx microstructures acts as a ionic conductor. Under the action of a negative/positive gate voltage, oxygen ions are driven towards/away from the Co/MOx interface, modifying its oxidation state and therefore the anisotropy of the magnetic stacks \cite{Manchon2008a}. Magneto-ionic effects, i.e. modification of the magnetic anisotropy driven by ion migration, generally reversible and non-volatile, have also been reported in other works \cite{Bi2014,Bauer2015,Zhou2016}.

Similarly, in the Pt/Co/Tb/Al stack studied here we expect that the application of a negative gate voltage results into the oxidation first of the top Al layer and consequently  of the Tb layer. This is confirmed by our hard x-ray photoelectron spectroscopy measurements, which reveal that the Al layer, which was only partially oxidised in the pristine sample (giving two peaks in the spectrum) becomes completely oxidised after the gating (only one peak characteristic of the oxide). A weak change appears in the spectrum of the Tb layer, while it keeps the features of metallic Tb (Fig.\ref{fig:1}(c-d)). The partial oxidation of the Tb layer should result in a decrease of the number of  magnetic Tb atoms, and therefore to an increase of the spontaneous magnetization $M_s$.

Note now that the Kerr contrast used for this investigation does not allow measuring quantitatively the effective magnetization in the gated area of the sample.
Also, no other quantitative probe can access to the local magnetization of layers buried several nanometers below the micron-size electrode surface.

For this reason, in order to further investigate the role of the voltage-driven  oxidation of the Tb layer on the magnetic parameters  and the DW dynamics, we have studied a series of reference Pt/Co/Tb/Pt multilayers, in which the ratio of the Tb and Co thicknesses was varied, resulting in a variety of magnetic parameters. The samples were grown with the same procedure used for the capacitor-like sample, but were studied in the form of continuous layers (see Methods). The spontaneous magnetization $M_{\mathrm{s}}$ and the anisotropy field, $\mu_0H_{\mathrm{k}}$ were measured by VSM-SQUID and the DW dyamics with polar MOKE microscopy as for the capacitor-like sample.

Fig. \ref{fig:2} (b) shows the variation of the spontaneous magnetization and of the anisotropy field measured as function of $t_{\mathrm{eff}}=\frac{t_{\mathrm{Tb}}}{t_{\mathrm{Co}}+t_{\mathrm{Tb}}}$, which we have chosen as the figure of merit reflecting the weight of the Tb layer in the magnetic stack. We note that as $t_{\mathrm{eff}}$ increases, the spontaneous magnetization decreases - as expected from the antiparallel alignment of Tb and Co magnetic moments -
and the effective anisotropy field increases. Remarkably, as the Tb content increases, the measured DMI field increases, consistently with the decrease of $M_s$ and the increase of $K_{eff}$ (Fig. \ref{fig:2} (c), Table I and Supplemental Material).

The variation of the magnetic parameters strongly modifies the features of the DW speed curves driven by an out-of-plane magnetic field. Figure \ref{fig:2}(d) shows the velocity curves measured for a selection of reference samples (called A, B and C in the following) whose magnetic parameters are shown in Table I. For  sample A, for which $t_{\mathrm{eff}}$ is the largest and $M_s$ the smallest, the DW velocity increases very slowly as a function of $B_{\mathrm{z}}$,  not reaching the depinning regime even for field $B_{\mathrm{z}}$=\SI{200}{\milli\tesla} where its velocity is only of $\approx$ \SI{50}{\meter/\second}. We attribute the large depinning field  $H_{dep} \propto ~\sigma/M_{\mathrm{s}}$ to the large DW energy density due to the large effective anisotropy and the low spontaneous magnetization.

In samples B and C, as $t_{\mathrm{eff}}$ decreases leading to the increase of $M_s$ and the decrease of $K_{eff}$, the depinning field decreases and the DW velocity curves show the features expected for two-dimensional chiral N\'{e}el DWs in a system with large DMI, \textit{i.e.} a large DW velocity that saturates  after the Walker field \cite{Pham2016,Krizakova2019,PenaGarcia2021}.

 \begin{table*}[t]
\caption[]{\textbf{Micromagnetic parameters measured for the reference samples and for the capacitor-like sample before and after gating. Slanted values are estimated from the micromagnetic simulations.}}
\begin{tabular}{lcccccccc} \hline \\
 Sample            && $t_{eff}$ & $M_{s}$ & $\mu_{0}H_{K}^{*}$ &  $K_{eff}$  &   $\mu_{0}H_{DMI}^{*}$ &    $D_{s}$ &     $A$ \\
                   & &          &[\si{\mega\ampere/\meter} ]&   [\si{\tesla}]       &  [\si{\mega\joule/\meter^3}]  &       [\si{\milli\tesla}]         & [\si{\pico\joule/\meter}]   &   [\si{\pico\joule/\meter}]           \\ \hline
\\
\textit{Reference samples} \\
(A) Pt/Co(0.7)/Tb(1.16) &&  0.62      &0.32       & 2.86   &    0.45   &    $>$350             &      $>$0.7  &   5                \\

(B) Pt/Co(0.9)/Tb(0.8)  &&  0.47      &  0.62       & 0.62   &    0.43   &   218               &    0.92   &   7    \\

(C) Pt/Co(1.1)/Tb(0.8) && 0.42         &  0.74   & 0.8  &    0.30    &   134               &    1      &   8  \\

 \\
\hline\\
\textit{Capacitor-like sample} \\
``\textit{state 0}'' &&  0.56    &0.42        & 1.7   &    0.24   &    150           &    \textit{0.5}  &   \textit{5}   \\
``\textit{state 1}''  & & \textit{0.5}      &  \textit{0.7}      & 0.3   &    \textit{0.12}   &   100          &    \textit{0.9}   &   \textit{8}    \\

``\textit{state 2}'' && \textit{0.47}         &  \textit{0.8}      & 0.28  &    \textit{0.11}    &   80               &  \textit{0.9}      &  \textit{10}  \\
\\

\hline

    \end{tabular}
\end{table*}

As shown in our previous work \cite{Pham2016,Krizakova2019}, by combining the measured Walker velocity ($v_W=\gamma \pi D/(2 M_s)$) and the value of the DMI field, it is possible to estimate the value of the exchange stiffness $A$,  for which the DMI constant $D$ extracted from the two measurements are the closest.  Values changing between \SI{15}{\pico\joule/\meter}  and  \SI{5}{\pico\joule/\meter}  are found, as the Tb content increases (Table I and Figure \ref{fig:2}(c) ). This decrease of $A$ is in line with the results reported for intermetallic amorphous alloys
\cite{Katayama1978} and in  GdCo thin films \citep{Caretta2018,Krizakova2019}.

The effective interfacial DMI constants, which are weakly dependent on the sample composition,  are shown in Table I for samples A, B and C.

The results of  micromagnetic simulations for samples B and C, using the measured magnetic parameters, show a good agreement with the experimental velocity curves, and can be found in the Supplementary Material.

\subsection{Tuning of the micromagnetic parameters in Pt/Co/Tb/Al after gating}

Let's now compare the DW velocity curves obtained for the capacitor-like Pt/Co/Tb/Al stack with those of the reference samples. The variation of the DW  velocities between "\textit{state 0}",  "\textit{state 1}" and "\textit{state 2}" follows the same pattern as those of samples A, B and C, where the spontaneous magnetization increases by decreasing the Tb content.

The slow increase of the DW velocity in the pristine state ("\textit{state 0}") is consistent with the measured low magnetization and the large magnetic anisotropy leading to a large depinning field, as observed in reference sample A.
The decrease of the depinning field and the strong increase of the DW velocity after gating reflect the increase of the spontaneous magnetization (like in samples B and C) and are coherent with the measured decrease of the in-plane saturation field. We can then confirm that the huge change  of the DW velocity after gating is due to the effect of the oxygen ion migration that leads to the partial oxidation of the Tb layer.

The missing magnetic parameters ($M_s$ and $A$) allowing us to quantitatively interpret the DW velocity curves after gating were obtained by carrying out a set of  micromagnetic simulations using the Mumax3 software \cite{Vansteenkiste2014}  (see Supplementary Material).  The simulations consisted in evaluating the DW speed at $B_{\mathrm{z}}=$ \SI{0.35}{\tesla} using the values of the experimental anisotropy and DMI fields, while varying the exchange stiffness and the spontaneous magnetization. The generated phase diagram allowed us to  determine the set of $A$ and $M_s$ values that provides the best agreement with the experimental velocities. This set of parameters is shown in Table I~. The results of the simulations and the details of the methodology are detailed in the Supplementary Material file.

Some interesting features emerge: an increase of $M_s$  (corresponding to a decrease of $\approx$ 0.3nm of the thickness of the metallic Tb layer, as extrapolated from the reference sample magnetic parameters) is observed after gating. In parallel, the interfacial DMI constant $D_s$ is observed to increase by a factor 2, going from $D_s \approx 0.5 pJ/m^2$ in the pristine state to $D_s \approx 0.9 pJ/m^2$ after gating. Since the oxidation of the Tb layer is expected to be inhomogeneous and to affect the Tb/Co interface, this result suggests that the DMI at the Co/TbOx interface is larger than that of the Co/Tb interface.

To conclude, we confirm that the huge variation of the field-driven domain wall velocities induced by the negative voltage is due to the partial oxidation of  the Tb layer, leading to an increase of the magnetization, a decrease of the anisotropy and a strong increase of the DMI constant in the region below the micropatterned electrodes.
The observed effect is reversible and non volatile, as the magnetic properties are kept for several weeks after the application of the gate voltage. Such magneto-ionic effect,  driven by the migration of oxygen ions across the ZrO$_2$  ionic conductor, may be easily extended to downscaled electrodes sizes \cite{Fassatoui2021} and its efficiency increased by optimizing the ionic conductor thickness.

The results of this work may have some promising application in low power logic devices and reprogrammable memories based on domain walls \cite{Raymenants2021}.

\section{Methods}
\subsection{Sample preparation}
Capacitor-like structures were prepared using the following procedure.  Ta(4)/Pt(4)/Co(1)/Tb(0.8)/Al(3) magnetic stacks (thicknesses in nm) were deposited by magnetron sputtering on Si/SiO$_2$ wafers.
After patterning the films into 1-20-$\mu$m wide strips by electron beam lithography (EBL) and ion-beam etching,  a 10 nm thick ZrO$_2$ dielectric layer was deposited by atomic layer deposition. The oxide layer, grown at 100\textdegree C, has an amorphous structure. Finally, 3 nm thick Pt electrodes (sizes 5-20 $\mu$m) acting as local gates, were patterned by EBL  and lift-off.

Before patterning, the magnetization of the sample was measured by VSM-SQUID magnetometry. The measurement of M$_S$ versus temperature shows the presence of a compensation temperature at around 100 K. The effective anisotropy energy is obtained from the measured in-plane saturation field.

 The Ta(4)/Pt(3)/Co(0.7-1.1)/Tb(0.8-1.52)/Pt(2) multilayers used as reference samples were grown with the same method as the previous sample. The magnetic properties were measured by VSM-SQUID.

\subsection{Domain wall dynamics}
Domain wall dynamics was measured by polar magneto-optical microscopy.
In the differential images reported in Figure \ref{fig:reproducibility} the white contrast corresponds to the displacement of the DWs driven by a 20~ms~long $B_z$=15mT magnetic field pulse obtained with an  electromagnet. For larger magnetic fields,  up to $B_{z}$=400~mT, 200 $\mu$m-diameter microcoils associated to a fast pulse current generator providing magnetic pulses down to 20~ns long with  rise/fall times of $\approx$ 5~ns were used.
The film magnetization was first saturated in the out-of-plane direction. An opposite  magnetic field pulse $B_{z}$ was then applied to nucleate a reverse domain. The DW velocity was obtained from the displacement of the domain wall during the application of further magnetic field pulses, divided by the total duration of the pulses. The domain wall velocity as a function of a static in-plane magnetic field was measured using a $B_z$=150mT as driving field.
The domain wall velocities were measured in the region below the patterned Pt electrodes before and after the application of the negative gate voltage, once the electrical leads have been removed. The gate voltage V$_g$=-8~V for 430~s was applied to obtain "\textit{state 1}",  V$_g$=-10~V for 600~s to obtain "\textit{state 2}".

\subsection{Hard X-ray Photoelectron Spectroscopy (HAXPES) measurements}
Hard X-ray Photoelectron Spectroscopy measurements were carried out at the GALAXIES beamline of the French synchrotron SOLEIL. The measurements were  carried out at room temperature with a photon energy of 9 keV which is a good compromise between penetration depth and photoionisation cross section. The Al 1s and Tb 4s spectra were repeated several times for a total acquisition time of around 30 minutes to improve S/N ratio.

\section{Acknowledgements}
 We acknowledge the support of the Agence Nationale de la Recherche (projects ANR-17-CE24-0025 (TOPSKY) and of the DARPA TEE program through Grant No. MIPR HR0011831554. The authors acknowledge funding from the European Union’s Horizon 2020 research and innovation program under Marie Sklodowska-Curie Grant Agreement No. 754303 and No. 860060 “Magnetism and the effect of Electric Field” (MagnEFi). J.P.G. also thanks the Laboratoire d\textquotesingle Excellence LANEF in Grenoble (ANR-10-LABX-0051) for its support. B. Fernandez, T. Crozes, Ph. David, E. Mossang and E. Wagner are acknowledged for their technical help.

\newpage

\section{Supplementary Material}

\subsection{Details of reference samples and domain wall velocity vs in-plane field}

The composition of the reference samples is shown in Table I. The Pt/Co/Tb multilayers are deposited by magnetron sputtering on Si/SiO$_2$ substrates with a 4~nm Ta buffer layer and a 2~nm Pt capping layer.

\begin{figure}[h!]
    \centering
    \includegraphics[scale=0.55]{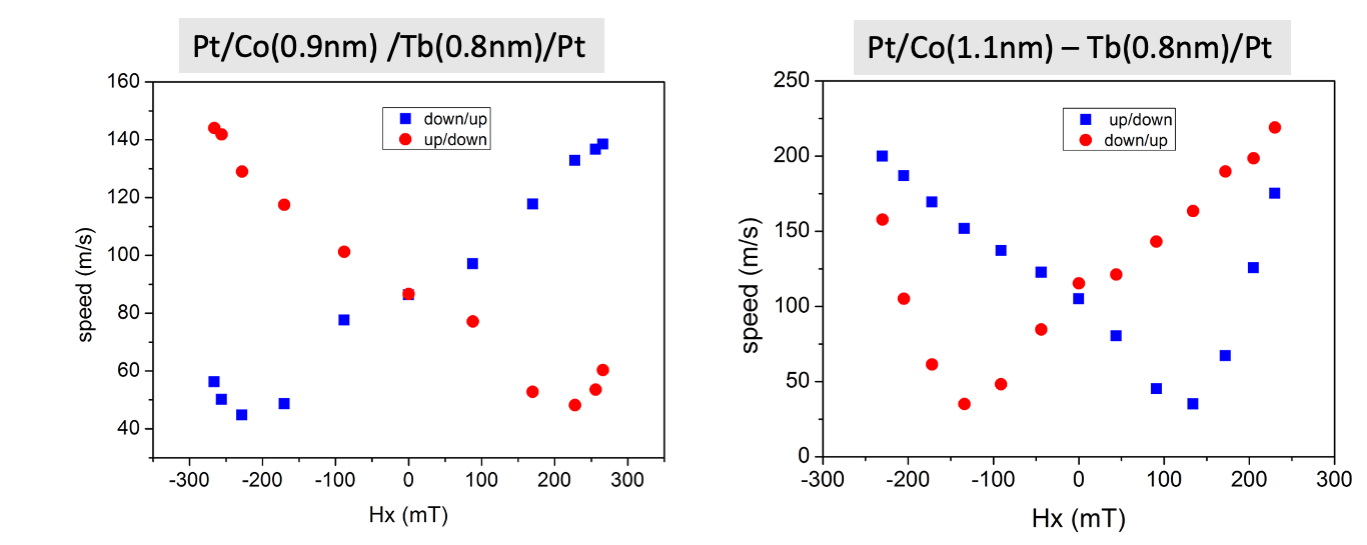}
    \caption{Domain wall velocity driven by an out-of-plane magnetic field ($B_z$=150 mT) measured as a function of a static in-plane field $B_x$ for reference samples B and C.}
    \label{fig:S1}
\end{figure}

The domain wall velocities in reference samples B and C, driven by the out-of-plane magnetic field $B_z$=150 mT, were measured as a function of the static in-plane magnetic field $B_x$. This measurement allows us to obtain the value of the DMI field, which corresponds with the field where the domain wall velocity vs. $B_x$ reaches a minimum.  The velocity curve measured for sample A did not reach a minimum for the maximum field available in our Kerr set-up.
The domain wall speed curves for samples B and C are shown in Figure \ref{fig:S1}.

\begin{table*}[b]
\caption[]{\textbf{Composition (thicknesses in nm) and magnetic parameters of the reference samples. $t_{eff}=({t_{Tb}}/{t_{Co}+t_{Tb}})$} is the figure of merit reflecting the weight of the Tb layer in the magnetic stack. }
\begin{tabular}{lcccccc}   \hline \\
 Reference sample            && $t_{eff}$ & $M_{s}$ & $\mu_{0}H_{K}^{*}$ &  $K_{eff}$      \\
                   & &          &[\si{\mega\ampere/\meter} ]&   [\si{\tesla}]       &  [\si{\mega\joule/\meter^3}]                \\ \hline
\\

 Pt(3)/Co(0.7)/Tb(1.16) $^{(A)}$ &&  0.62      &0.32       & 2.86   &    0.45                                   \\
 Pt(3)/Co(0.7)/Tb(0.8) &&  0.53      &0.53       & 1.77   &    0.47                   \\

Pt(3)/Co(0.9)/Tb(0.8) $^{(B)}$  &&  0.47      &  0.62       & 0.62   &    0.43                      \\

Pt(3)/Co(1.1)/Tb(0.8) $^{(C)}$ && 0.42         &  0.74   & 0.8  &    0.30                      \\
Pt(3)/Co(1.3)/Tb(0.8) && 0.38         &  0.85   & 0.39  &    0.17                     \\

 \\

    \end{tabular}
\end{table*}

 \subsection{Micromagnetic simulations, methods}
 Micromagnetic simulations are performed with the  Mumax3 software \cite{Vansteenkiste2014} which solves numerically the Landau-Lifshitz-Gilbert equation using a finite-difference approach. Note that the Co/Tb  bilayer in the Pt/Co/Tb stacks may be modeled as an effective ferromagnetic layer due to the fact that the magnetization at room temperature  is far from the angular and magnetic compensation point  \cite{Haltz2021}. Therefore, the Co/Tb bilayer may be considered as a single ferromagnetic layer, with  effective micromagnetic parameters.

  The experimentally measured fields, $\mu_0H_{\mathrm{DMI}}, \mu_0H_{\mathrm{K}}$, are used as inputs for both the reference and gating samples. The spontaneous magnetization could be measured with VSM-SQUID only for the reference samples and for the capacitor-like sample in the pristine state.

 For the reference samples, the exchange stiffness $A$ was deduced combining the plateau DW velocity and the DMI effective field \cite{Krizakova2019,PenaGarcia2021}, leading to the relationship:
\begin{equation}
    A=2\mu_0H_{\mathrm{K}}M_{\mathrm{s}}\left(\frac{v_{\mathrm{plateau}}}{\pi\gamma\mu_0H_{\mathrm{DMI}}}\right)^2
\end{equation}

 \subsection{Micromagnetic simulations of reference samples}

 A Néel domain wall is initialised in the center of a  \SI{1}{\micro\meter^2} strip, which is relaxed to its ground state. The thickness of the strip is the sum of the Co and Tb layers, $t_{\mathrm{tot}}=t_{\mathrm{Co}}+t_{\mathrm{Tb}}$. The strip is discretised in a grid of $1024\times1024\times1$ cells.  The Néel DW is driven by an out-of-plane magnetic field within a \SI{1}{\micro\meter^2} moving-frame window, so as to keep the domain wall in its center.

 \begin{figure}[t!]
    \centering
    \includegraphics[scale=0.7]{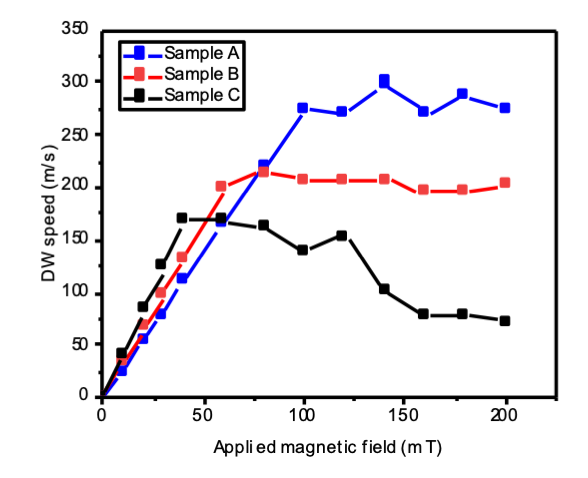}
    \caption{\textbf{Simulated field-driven DW velocity for the reference samples.} The micromagnetic parameters are those of Table 1 of the main text. }
   \label{fig:S2}
\end{figure}

The damping parameter is set to 0.2 as is the maximum value that allows to fulfill the Walker field condition, $\mu_0H_{\mathrm{W}}\approx\alpha\frac{\pi}{2}\mu_0H_{\mathrm{DMI}}$ for sample B and C.  No defects or temperature are considered, as the plateau velocity is mostly independent on defects and temperature \cite{Krizakova2019}. Figure \ref{fig:S2} displays the micromagnetic simulations of the reference samples, showing a remarkably good agreement  with the experimental data of samples B and C. On the other hand, the simulated plateau DW velocity for sample A, for which $M_s$ is the weakest among the three samples, appears to be larger than for samples B and C. These simulations do not take into account the presence of strong pinning in the real samples, that prevent the DW to enter into the flow regime, as observed experimentally.

\subsection{Micromagnetic simulations for the capacitor-like sample }

The micromagnetic simulations of the domain wall dynamics for the magnetic states induced after gating "(\textit{state 1}" and "\textit{state 2}")  are more complex due to the impossibility to measure locally the spontaneous magnetization. For experimental field much larger than the Walker field the DW is precessing, and its averaged velocity is given by:

{\begin{equation}
    <v>=\dfrac{\gamma_0\Delta H_z}{\alpha}-\dfrac{\Delta}{\alpha}\dfrac{2\pi}{T}
\end{equation}}

where $T$ is the precession period
\begin{equation}
    T=\dfrac{1+\alpha^2}{\gamma_0}\int_0^{2\pi}\dfrac{d\varphi(t)}{H_z-\alpha\sin{\varphi(t)}(\frac{\pi}{2}H_{\mathrm{DMI}}-H_{\mathrm{DW}}\cos{\varphi(t)})}
\end{equation}

Therefore, the DW speed depends on the value of the exchange stiffness, $A$, the spontaneous magnetization, $M_s$ and the damping parameter, $\alpha$.

\begin{figure}[t!]
    \centering
    \includegraphics[scale=0.55]{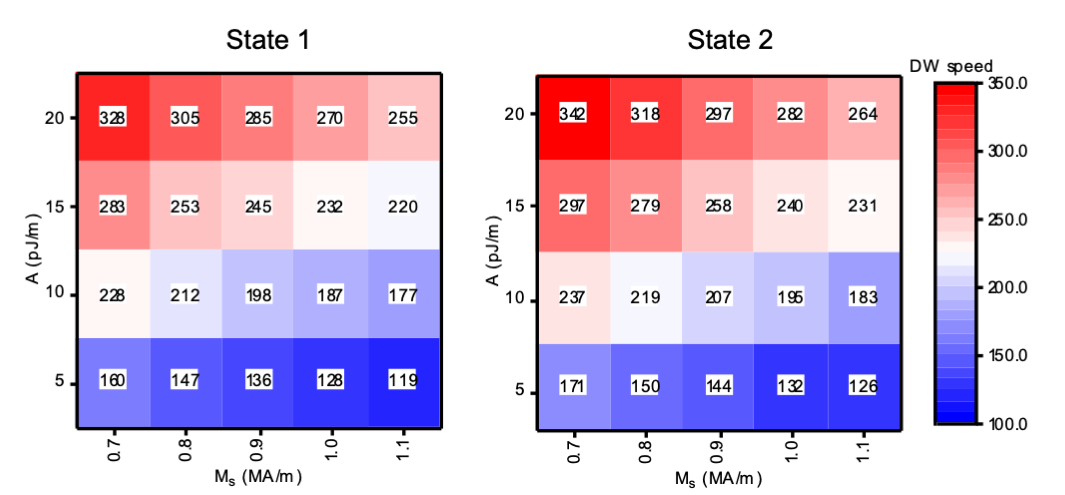}
    \caption{\textbf{$\mathbf{M_s-A}$ phase diagram showing the domain wall velocity} obtained for a field $B_z$=\SI{300}{\milli\tesla}for a combination of exchange stiffness and spontaneous magnetization values, for the two different magnetic states after gating. The magnetic parameters correspond with the Pt/Co/Tb/Ta stack after gating ("\textit{state 1").}}
   \label{fig:S3}
\end{figure}

\begin{figure}[t!]
    \centering
    \includegraphics[scale=0.45]{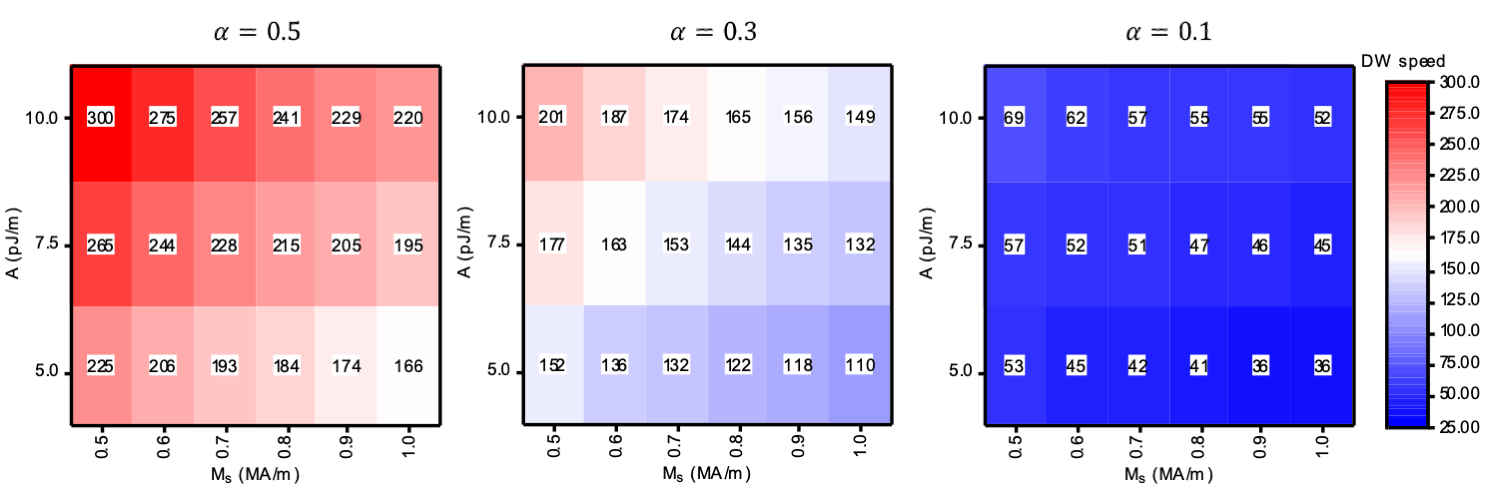}
    \caption{\textbf{$\mathbf{M_s-A}$ phase diagram showing the domain wall velocity} obtained for a field $B_z$=300 mT for a combination of exchange stiffness and spontaneous magnetization values, for a damping parameter between 0.1 and 0.5. The magnetic parameters correspond with the Pt/Co/Tb/Ta stak after gating ("\textit{state 1").}}
   \label{fig:S4}
\end{figure}

We performed a set of micromagnetic simulations  for different values of exchange  and spontaneous magnetization for $\mu_0H_{\mathrm{z}}$= \SI{0.3}{\tesla} which is the  field for which in "\textit{state 1}" and "\textit{state 2}" after gating the domain walls enter into the \textit{flow} regime. For this simulations we set the damping parameter to 0.5. This leads to a $M_s-A$ phase diagram (Figure \ref{fig:S3}). The best combination of $M_s$ and $A$ which reproduces the experimental speed is chosen as the magnetic parameters for each state. The role of the damping parameter can be observed in the phase diagram of Figure \ref{fig:S4}, where the $M_s-A$ phase diagram was generated for different values of damping, for "\textit{state 1}". As it can be observed, neither $\alpha=0.1$ nor $ 0.3$ allow us to reproduce the DW velocities experimentally observed for a combination of exchange and spontaneous magnetization, within the range determined from the reference samples. A similar behaviour is observed for "\textit{state 2}". Therefore, we set the damping parameter equal to 0.5.
Finally, the experimental curve were reproduced by taking into account  disorder and  by dividing the sample into grains using the Voronoi tessellation \cite{Vansteenkiste2014,leliaert2014current}. In each grain the micromagnetic parameters  change in a correlated way. We have set the grain size to 15nm and the standard deviation of 10\%.

\begin{figure}[h!]
    \centering
    \includegraphics[scale=0.60]{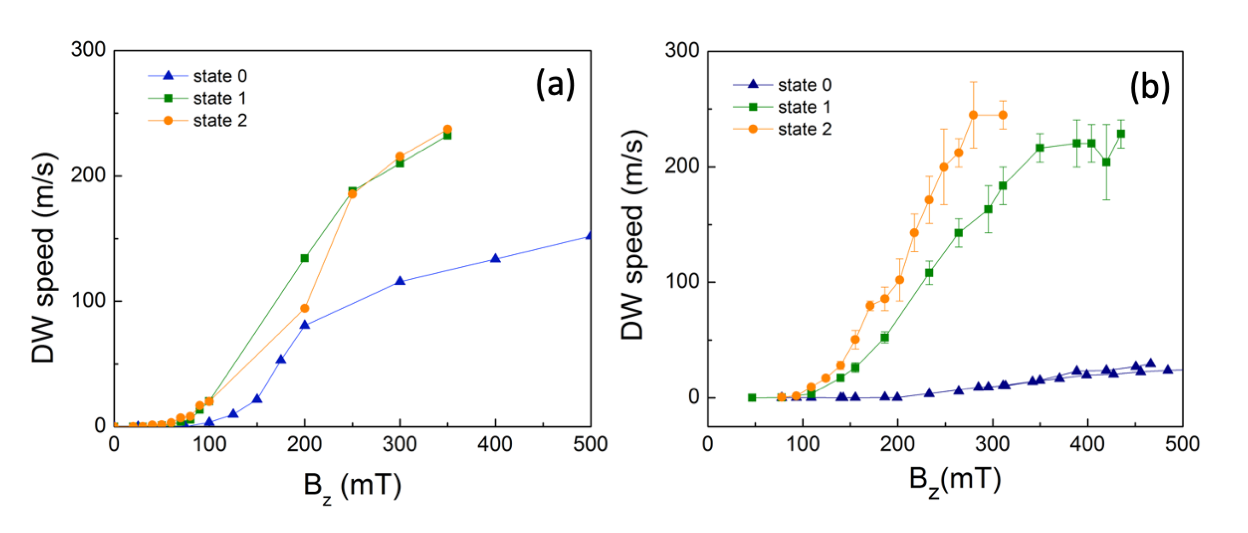}
    \caption{\textbf{Comparison of experimental and simulated DW velocity in the capacitor-like Pt/Co/Tb/Ta sample before ("\textit{state 0}") and after gating ("\textit{state 1}" and "\textit{state 2 }"):} (a):  experiments; (b) micromagnetic simulations.}
    \label{fig:S5}
\end{figure}

The results of the simulations giving rise to the best fit to the experimental data are shown in Figure \ref{fig:S5} where they are compared with the experimental data.

\begin{thebibliography}{29}
\expandafter\ifx\csname natexlab\endcsname\relax\def\natexlab#1{#1}\fi
\expandafter\ifx\csname bibnamefont\endcsname\relax
  \def\bibnamefont#1{#1}\fi
\expandafter\ifx\csname bibfnamefont\endcsname\relax
  \def\bibfnamefont#1{#1}\fi
\expandafter\ifx\csname citenamefont\endcsname\relax
  \def\citenamefont#1{#1}\fi
\expandafter\ifx\csname url\endcsname\relax
  \def\url#1{\texttt{#1}}\fi
\expandafter\ifx\csname urlprefix\endcsname\relax\def\urlprefix{URL }\fi
\providecommand{\bibinfo}[2]{#2}
\providecommand{\eprint}[2][]{\url{#2}}

\bibitem[{\citenamefont{Thiaville et~al.}(2012)\citenamefont{Thiaville, Rohart,
  Ju{\'e}, Cros, and Fert}}]{Thiaville2012}
\bibinfo{author}{\bibfnamefont{A.}~\bibnamefont{Thiaville}},
  \bibinfo{author}{\bibfnamefont{S.}~\bibnamefont{Rohart}},
  \bibinfo{author}{\bibfnamefont{{\'E}.}~\bibnamefont{Ju{\'e}}},
  \bibinfo{author}{\bibfnamefont{V.}~\bibnamefont{Cros}}, \bibnamefont{and}
  \bibinfo{author}{\bibfnamefont{A.}~\bibnamefont{Fert}}, \bibinfo{journal}{EPL
  (Europhysics Letters)} \textbf{\bibinfo{volume}{100}}, \bibinfo{pages}{57002}
  (\bibinfo{year}{2012}).

\bibitem[{\citenamefont{Miron et~al.}(2010)\citenamefont{Miron, Gaudin,
  Auffret, Rodmacq, Schuhl, Pizzini, Vogel, and Gambardella}}]{Miron2010}
\bibinfo{author}{\bibfnamefont{I.~M.} \bibnamefont{Miron}},
  \bibinfo{author}{\bibfnamefont{G.}~\bibnamefont{Gaudin}},
  \bibinfo{author}{\bibfnamefont{S.}~\bibnamefont{Auffret}},
  \bibinfo{author}{\bibfnamefont{B.}~\bibnamefont{Rodmacq}},
  \bibinfo{author}{\bibfnamefont{A.}~\bibnamefont{Schuhl}},
  \bibinfo{author}{\bibfnamefont{S.}~\bibnamefont{Pizzini}},
  \bibinfo{author}{\bibfnamefont{J.}~\bibnamefont{Vogel}}, \bibnamefont{and}
  \bibinfo{author}{\bibfnamefont{P.}~\bibnamefont{Gambardella}},
  \bibinfo{journal}{Nat. Mater.} \textbf{\bibinfo{volume}{9}},
  \bibinfo{pages}{230} (\bibinfo{year}{2010}).

\bibitem[{\citenamefont{Miron et~al.}(2011)\citenamefont{Miron, Moore,
  Szambolics, Buda-Prejbeanu, Auffret, Rodmacq, Vogel, Bonfim, Schuhl, and
  Gaudin}}]{Miron2011}
\bibinfo{author}{\bibfnamefont{I.~M.} \bibnamefont{Miron}},
  \bibinfo{author}{\bibfnamefont{T.}~\bibnamefont{Moore}},
  \bibinfo{author}{\bibfnamefont{H.}~\bibnamefont{Szambolics}},
  \bibinfo{author}{\bibfnamefont{L.~D.} \bibnamefont{Buda-Prejbeanu}},
  \bibinfo{author}{\bibfnamefont{S.}~\bibnamefont{Auffret}},
  \bibinfo{author}{\bibfnamefont{S.}~\bibnamefont{Rodmacq},
  \bibfnamefont{B.and~Pizzini}},
  \bibinfo{author}{\bibfnamefont{J.}~\bibnamefont{Vogel}},
  \bibinfo{author}{\bibfnamefont{M.}~\bibnamefont{Bonfim}},
  \bibinfo{author}{\bibfnamefont{A.}~\bibnamefont{Schuhl}}, \bibnamefont{and}
  \bibinfo{author}{\bibfnamefont{G.}~\bibnamefont{Gaudin}},
  \bibinfo{journal}{Nat. Mater.} \textbf{\bibinfo{volume}{10}},
  \bibinfo{pages}{419} (\bibinfo{year}{2011}).

\bibitem[{\citenamefont{Ryu et~al.}(2013)\citenamefont{Ryu, L., Yang, and
  Parkin}}]{Ryu2013}
\bibinfo{author}{\bibfnamefont{K.-S.} \bibnamefont{Ryu}},
  \bibinfo{author}{\bibfnamefont{T.}~\bibnamefont{L.}},
  \bibinfo{author}{\bibfnamefont{S.-H.} \bibnamefont{Yang}}, \bibnamefont{and}
  \bibinfo{author}{\bibfnamefont{S.}~\bibnamefont{Parkin}},
  \bibinfo{journal}{Nat. Nanotechnol.} \textbf{\bibinfo{volume}{8}},
  \bibinfo{pages}{527} (\bibinfo{year}{2013}).

\bibitem[{\citenamefont{Emori et~al.}(2013)\citenamefont{Emori, Bauer, Ahn,
  Martinez, and Beach}}]{Emori2013}
\bibinfo{author}{\bibfnamefont{S.}~\bibnamefont{Emori}},
  \bibinfo{author}{\bibfnamefont{U.}~\bibnamefont{Bauer}},
  \bibinfo{author}{\bibfnamefont{S.-M.} \bibnamefont{Ahn}},
  \bibinfo{author}{\bibfnamefont{E.}~\bibnamefont{Martinez}}, \bibnamefont{and}
  \bibinfo{author}{\bibfnamefont{G.}~\bibnamefont{Beach}},
  \bibinfo{journal}{Nat. Mater.} \textbf{\bibinfo{volume}{12}},
  \bibinfo{pages}{611} (\bibinfo{year}{2013}).

\bibitem[{\citenamefont{Pham et~al.}({2016})\citenamefont{Pham, Vogel, Sampaio,
  Vanatka, Rojas-Sanchez, Bonfim, Chaves, Choueikani, Ohresser, Otero
  et~al.}}]{Pham2016}
\bibinfo{author}{\bibfnamefont{T.~H.} \bibnamefont{Pham}},
  \bibinfo{author}{\bibfnamefont{J.}~\bibnamefont{Vogel}},
  \bibinfo{author}{\bibfnamefont{J.}~\bibnamefont{Sampaio}},
  \bibinfo{author}{\bibfnamefont{M.}~\bibnamefont{Vanatka}},
  \bibinfo{author}{\bibfnamefont{J.-C.} \bibnamefont{Rojas-Sanchez}},
  \bibinfo{author}{\bibfnamefont{M.}~\bibnamefont{Bonfim}},
  \bibinfo{author}{\bibfnamefont{D.~S.} \bibnamefont{Chaves}},
  \bibinfo{author}{\bibfnamefont{F.}~\bibnamefont{Choueikani}},
  \bibinfo{author}{\bibfnamefont{P.}~\bibnamefont{Ohresser}},
  \bibinfo{author}{\bibfnamefont{E.}~\bibnamefont{Otero}},
  \bibnamefont{et~al.}, \bibinfo{journal}{{EPL}}
  \textbf{\bibinfo{volume}{{113}}}, \bibinfo{pages}{67001}
  (\bibinfo{year}{{2016}}).

\bibitem[{\citenamefont{Schellekens et~al.}(2012)\citenamefont{Schellekens,
  van~den Brink, Franken, Swagten, and Koopmans}}]{Schellekens2012}
\bibinfo{author}{\bibfnamefont{A.}~\bibnamefont{Schellekens}},
  \bibinfo{author}{\bibfnamefont{A.}~\bibnamefont{van~den Brink}},
  \bibinfo{author}{\bibfnamefont{J.}~\bibnamefont{Franken}},
  \bibinfo{author}{\bibfnamefont{H.}~\bibnamefont{Swagten}}, \bibnamefont{and}
  \bibinfo{author}{\bibfnamefont{B.}~\bibnamefont{Koopmans}},
  \bibinfo{journal}{Nat. Commun.} \textbf{\bibinfo{volume}{3}},
  \bibinfo{pages}{847} (\bibinfo{year}{2012}).

\bibitem[{\citenamefont{Bauer et~al.}(2012)\citenamefont{Bauer, Emori, and
  Beach}}]{Bauer2012a}
\bibinfo{author}{\bibfnamefont{U.}~\bibnamefont{Bauer}},
  \bibinfo{author}{\bibfnamefont{S.}~\bibnamefont{Emori}}, \bibnamefont{and}
  \bibinfo{author}{\bibfnamefont{G.~S.~D.} \bibnamefont{Beach}},
  \bibinfo{journal}{Appl. Phys. Lett.} \textbf{\bibinfo{volume}{101}},
  \bibinfo{pages}{172403} (\bibinfo{year}{2012}).

\bibitem[{\citenamefont{Bernand-Mantel
  et~al.}(2013)\citenamefont{Bernand-Mantel, Herrera-Diez, Ranno, Pizzini,
  Vogel, Givord, Auffret, Boulle, Miron, and Gaudin}}]{Bernand-Mantel2013}
\bibinfo{author}{\bibfnamefont{A.}~\bibnamefont{Bernand-Mantel}},
  \bibinfo{author}{\bibfnamefont{L.}~\bibnamefont{Herrera-Diez}},
  \bibinfo{author}{\bibfnamefont{L.}~\bibnamefont{Ranno}},
  \bibinfo{author}{\bibfnamefont{S.}~\bibnamefont{Pizzini}},
  \bibinfo{author}{\bibfnamefont{J.}~\bibnamefont{Vogel}},
  \bibinfo{author}{\bibfnamefont{D.}~\bibnamefont{Givord}},
  \bibinfo{author}{\bibfnamefont{S.}~\bibnamefont{Auffret}},
  \bibinfo{author}{\bibfnamefont{O.}~\bibnamefont{Boulle}},
  \bibinfo{author}{\bibfnamefont{I.~M.} \bibnamefont{Miron}}, \bibnamefont{and}
  \bibinfo{author}{\bibfnamefont{G.}~\bibnamefont{Gaudin}},
  \bibinfo{journal}{Appl. Phys. Lett.} \textbf{\bibinfo{volume}{102}},
  \bibinfo{pages}{122406} (\bibinfo{year}{2013}).

\bibitem[{\citenamefont{Chiba et~al.}(2012)\citenamefont{Chiba, Kawaguchi,
  Fukami, Ishiwata, Shimamura, and Ono}}]{Chiba2012}
\bibinfo{author}{\bibfnamefont{D.}~\bibnamefont{Chiba}},
  \bibinfo{author}{\bibfnamefont{M.}~\bibnamefont{Kawaguchi}},
  \bibinfo{author}{\bibfnamefont{S.}~\bibnamefont{Fukami}},
  \bibinfo{author}{\bibfnamefont{N.}~\bibnamefont{Ishiwata}},
  \bibinfo{author}{\bibfnamefont{K.}~\bibnamefont{Shimamura},
  \bibfnamefont{K.and~Kobayashi}}, \bibnamefont{and}
  \bibinfo{author}{\bibfnamefont{T.}~\bibnamefont{Ono}},
  \bibinfo{journal}{Nat. Commun.} \textbf{\bibinfo{volume}{3}},
  \bibinfo{pages}{888} (\bibinfo{year}{2012}).

\bibitem[{\citenamefont{Franke et~al.}(2015)\citenamefont{Franke, Van~de Wiele,
  Shirahata, H\"am\"al\"ainen, Taniyama, and van Dijken}}]{Franke2015}
\bibinfo{author}{\bibfnamefont{K.~J.~A.} \bibnamefont{Franke}},
  \bibinfo{author}{\bibfnamefont{B.}~\bibnamefont{Van~de Wiele}},
  \bibinfo{author}{\bibfnamefont{Y.}~\bibnamefont{Shirahata}},
  \bibinfo{author}{\bibfnamefont{S.~J.} \bibnamefont{H\"am\"al\"ainen}},
  \bibinfo{author}{\bibfnamefont{T.}~\bibnamefont{Taniyama}}, \bibnamefont{and}
  \bibinfo{author}{\bibfnamefont{S.}~\bibnamefont{van Dijken}},
  \bibinfo{journal}{Phys. Rev. X} \textbf{\bibinfo{volume}{5}},
  \bibinfo{pages}{011010} (\bibinfo{year}{2015}).

\bibitem[{\citenamefont{Koyama et~al.}(2018)\citenamefont{Koyama, Nakatani,
  Ieda, and Chiba}}]{Koyama2018}
\bibinfo{author}{\bibfnamefont{T.}~\bibnamefont{Koyama}},
  \bibinfo{author}{\bibfnamefont{Y.}~\bibnamefont{Nakatani}},
  \bibinfo{author}{\bibfnamefont{J.}~\bibnamefont{Ieda}}, \bibnamefont{and}
  \bibinfo{author}{\bibfnamefont{D.}~\bibnamefont{Chiba}},
  \bibinfo{journal}{Science Advances} \textbf{\bibinfo{volume}{4}},
  \bibinfo{pages}{eaav0265} (\bibinfo{year}{2018}).

\bibitem[{\citenamefont{Bauer et~al.}(2013)\citenamefont{Bauer, Emori, and
  Beach}}]{Bauer2013}
\bibinfo{author}{\bibfnamefont{U.}~\bibnamefont{Bauer}},
  \bibinfo{author}{\bibfnamefont{S.}~\bibnamefont{Emori}}, \bibnamefont{and}
  \bibinfo{author}{\bibfnamefont{G.~S.~D.} \bibnamefont{Beach}},
  \bibinfo{journal}{Nat. Nanotechnol.} \textbf{\bibinfo{volume}{8}},
  \bibinfo{pages}{411} (\bibinfo{year}{2013}).

\bibitem[{\citenamefont{Bi et~al.}(2014)\citenamefont{Bi, Liu, Newhouse-Illige,
  Xu, Rosales, Freeland, Mryasov, Zhang, Velthuis, and Wang}}]{Bi2014}
\bibinfo{author}{\bibfnamefont{C.}~\bibnamefont{Bi}},
  \bibinfo{author}{\bibfnamefont{Y.}~\bibnamefont{Liu}},
  \bibinfo{author}{\bibfnamefont{T.}~\bibnamefont{Newhouse-Illige}},
  \bibinfo{author}{\bibfnamefont{M.}~\bibnamefont{Xu}},
  \bibinfo{author}{\bibfnamefont{M.}~\bibnamefont{Rosales}},
  \bibinfo{author}{\bibfnamefont{J.~W.} \bibnamefont{Freeland}},
  \bibinfo{author}{\bibfnamefont{O.}~\bibnamefont{Mryasov}},
  \bibinfo{author}{\bibfnamefont{S.}~\bibnamefont{Zhang}},
  \bibinfo{author}{\bibfnamefont{S.~G.~E.} \bibnamefont{Velthuis}},
  \bibnamefont{and} \bibinfo{author}{\bibfnamefont{W.~G.} \bibnamefont{Wang}},
  \bibinfo{journal}{Phys. Rev. Lett.} \textbf{\bibinfo{volume}{113}},
  \bibinfo{pages}{267202} (\bibinfo{year}{2014}).

\bibitem[{\citenamefont{Bauer et~al.}(2015)\citenamefont{Bauer, Lide, Aik,
  Agrawal, Emori, Tuller, van Dijken, and Beach}}]{Bauer2015}
\bibinfo{author}{\bibfnamefont{U.}~\bibnamefont{Bauer}},
  \bibinfo{author}{\bibfnamefont{Y.}~\bibnamefont{Lide}},
  \bibinfo{author}{\bibfnamefont{J.~T.} \bibnamefont{Aik}},
  \bibinfo{author}{\bibfnamefont{P.}~\bibnamefont{Agrawal}},
  \bibinfo{author}{\bibfnamefont{S.}~\bibnamefont{Emori}},
  \bibinfo{author}{\bibfnamefont{H.}~\bibnamefont{Tuller}},
  \bibinfo{author}{\bibfnamefont{S.}~\bibnamefont{van Dijken}},
  \bibnamefont{and} \bibinfo{author}{\bibfnamefont{G.}~\bibnamefont{Beach}},
  \bibinfo{journal}{Nat. Mater.} \textbf{\bibinfo{volume}{14}},
  \bibinfo{pages}{174} (\bibinfo{year}{2015}).

\bibitem[{\citenamefont{Zhou et~al.}(2016)\citenamefont{Zhou, Yan, Jiang, Cui,
  Pan, and Song}}]{Zhou2016}
\bibinfo{author}{\bibfnamefont{X.}~\bibnamefont{Zhou}},
  \bibinfo{author}{\bibfnamefont{Y.}~\bibnamefont{Yan}},
  \bibinfo{author}{\bibfnamefont{M.}~\bibnamefont{Jiang}},
  \bibinfo{author}{\bibfnamefont{B.}~\bibnamefont{Cui}},
  \bibinfo{author}{\bibfnamefont{F.}~\bibnamefont{Pan}}, \bibnamefont{and}
  \bibinfo{author}{\bibfnamefont{C.}~\bibnamefont{Song}}, \bibinfo{journal}{The
  Journal of Physical Chemistry C} \textbf{\bibinfo{volume}{120}},
  \bibinfo{pages}{1633} (\bibinfo{year}{2016}).

\bibitem[{\citenamefont{Fassatoui et~al.}(2020)\citenamefont{Fassatoui, Garcia,
  Ranno, Vogel, Bernand-Mantel, B\'ea, Pizzini, and Pizzini}}]{Fassatoui2020}
\bibinfo{author}{\bibfnamefont{A.}~\bibnamefont{Fassatoui}},
  \bibinfo{author}{\bibfnamefont{J.} \bibnamefont{Pe\~{n}a Garcia}},
  \bibinfo{author}{\bibfnamefont{L.}~\bibnamefont{Ranno}},
  \bibinfo{author}{\bibfnamefont{J.}~\bibnamefont{Vogel}},
  \bibinfo{author}{\bibfnamefont{A.}~\bibnamefont{Bernand-Mantel}},
  \bibinfo{author}{\bibfnamefont{H.}~\bibnamefont{B\'ea}},
  \bibinfo{author}{\bibfnamefont{S.}~\bibnamefont{Pizzini}}, \bibnamefont{and}
  \bibinfo{author}{\bibfnamefont{S.}~\bibnamefont{Pizzini}},
  \bibinfo{journal}{Phys. Rev. Applied} \textbf{\bibinfo{volume}{14}},
  \bibinfo{pages}{064041} (\bibinfo{year}{2020}).

\bibitem[{\citenamefont{Fassatoui et~al.}(2021)\citenamefont{Fassatoui, Ranno,
  Peña~Garcia, Balan, Vogel, Béa, and Pizzini}}]{Fassatoui2021}
\bibinfo{author}{\bibfnamefont{A.}~\bibnamefont{Fassatoui}},
  \bibinfo{author}{\bibfnamefont{L.}~\bibnamefont{Ranno}},
  \bibinfo{author}{\bibfnamefont{J.}~\bibnamefont{Peña~Garcia}},
  \bibinfo{author}{\bibfnamefont{C.}~\bibnamefont{Balan}},
  \bibinfo{author}{\bibfnamefont{J.}~\bibnamefont{Vogel}},
  \bibinfo{author}{\bibfnamefont{H.}~\bibnamefont{Béa}}, \bibnamefont{and}
  \bibinfo{author}{\bibfnamefont{S.}~\bibnamefont{Pizzini}},
  \bibinfo{journal}{Small} \textbf{\bibinfo{volume}{17}},
  \bibinfo{pages}{2102427} (\bibinfo{year}{2021}).

\bibitem[{\citenamefont{Lin et~al.}(2016)\citenamefont{Lin, Vernier, Agnus,
  Garcia, Ocker, Zhao, Fullerton, and Ravelosona}}]{Lin2016}
\bibinfo{author}{\bibfnamefont{W.}~\bibnamefont{Lin}},
  \bibinfo{author}{\bibfnamefont{N.}~\bibnamefont{Vernier}},
  \bibinfo{author}{\bibfnamefont{G.}~\bibnamefont{Agnus}},
  \bibinfo{author}{\bibfnamefont{K.}~\bibnamefont{Garcia}},
  \bibinfo{author}{\bibfnamefont{B.}~\bibnamefont{Ocker}},
  \bibinfo{author}{\bibfnamefont{W.}~\bibnamefont{Zhao}},
  \bibinfo{author}{\bibfnamefont{E.~E.} \bibnamefont{Fullerton}},
  \bibnamefont{and}
  \bibinfo{author}{\bibfnamefont{D.}~\bibnamefont{Ravelosona}},
  \bibinfo{journal}{Nat. Commun.} \textbf{\bibinfo{volume}{7}},
  \bibinfo{pages}{13532} (\bibinfo{year}{2016}).

\bibitem[{\citenamefont{Guan et~al.}(2021)\citenamefont{Guan, Zhou, Li, Ma,
  Yang, and Parkin}}]{Guan2021}
\bibinfo{author}{\bibfnamefont{Y.}~\bibnamefont{Guan}},
  \bibinfo{author}{\bibfnamefont{X.}~\bibnamefont{Zhou}},
  \bibinfo{author}{\bibfnamefont{F.}~\bibnamefont{Li}},
  \bibinfo{author}{\bibfnamefont{T.}~\bibnamefont{Ma}},
  \bibinfo{author}{\bibfnamefont{S.-H.} \bibnamefont{Yang}}, \bibnamefont{and}
  \bibinfo{author}{\bibfnamefont{S.~S.~P.} \bibnamefont{Parkin}},
  \bibinfo{journal}{Nat. Commun.} \textbf{\bibinfo{volume}{12}},
  \bibinfo{pages}{5002} (\bibinfo{year}{2021}).

\bibitem[{\citenamefont{Je et~al.}(2013)\citenamefont{Je, Kim, Yoo, Min, Lee,
  and Choe}}]{Je2013}
\bibinfo{author}{\bibfnamefont{S.-G.} \bibnamefont{Je}},
  \bibinfo{author}{\bibfnamefont{D.-H.} \bibnamefont{Kim}},
  \bibinfo{author}{\bibfnamefont{S.-C.} \bibnamefont{Yoo}},
  \bibinfo{author}{\bibfnamefont{B.-C.} \bibnamefont{Min}},
  \bibinfo{author}{\bibfnamefont{K.-J.} \bibnamefont{Lee}}, \bibnamefont{and}
  \bibinfo{author}{\bibfnamefont{S.-B.} \bibnamefont{Choe}},
  \bibinfo{journal}{Phys. Rev. B} \textbf{\bibinfo{volume}{88}},
  \bibinfo{pages}{214401} (\bibinfo{year}{2013}).

\bibitem[{\citenamefont{Jeudy et~al.}(2018)\citenamefont{Jeudy,
  D\'{\i}az~Pardo, Savero~Torres, Bustingorry, and Kolton}}]{Jeudy2018}
\bibinfo{author}{\bibfnamefont{V.}~\bibnamefont{Jeudy}},
  \bibinfo{author}{\bibfnamefont{R.}~\bibnamefont{D\'{\i}az~Pardo}},
  \bibinfo{author}{\bibfnamefont{W.}~\bibnamefont{Savero~Torres}},
  \bibinfo{author}{\bibfnamefont{S.}~\bibnamefont{Bustingorry}},
  \bibnamefont{and} \bibinfo{author}{\bibfnamefont{A.~B.}
  \bibnamefont{Kolton}}, \bibinfo{journal}{Phys. Rev. B}
  \textbf{\bibinfo{volume}{98}}, \bibinfo{pages}{054406}
  (\bibinfo{year}{2018}).

\bibitem[{\citenamefont{Manchon et~al.}(2008)\citenamefont{Manchon, Ducruet,
  Lombard, Auffret, Rodmacq, Dieny, Pizzini, Vogel, Uhl\'{\i}\u{r},
  Hochstrasser et~al.}}]{Manchon2008a}
\bibinfo{author}{\bibfnamefont{A.}~\bibnamefont{Manchon}},
  \bibinfo{author}{\bibfnamefont{C.}~\bibnamefont{Ducruet}},
  \bibinfo{author}{\bibfnamefont{L.}~\bibnamefont{Lombard}},
  \bibinfo{author}{\bibfnamefont{S.}~\bibnamefont{Auffret}},
  \bibinfo{author}{\bibfnamefont{B.}~\bibnamefont{Rodmacq}},
  \bibinfo{author}{\bibfnamefont{B.}~\bibnamefont{Dieny}},
  \bibinfo{author}{\bibfnamefont{S.}~\bibnamefont{Pizzini}},
  \bibinfo{author}{\bibfnamefont{J.}~\bibnamefont{Vogel}},
  \bibinfo{author}{\bibfnamefont{V.}~\bibnamefont{Uhl\'{\i}\u{r}}},
  \bibinfo{author}{\bibfnamefont{M.}~\bibnamefont{Hochstrasser}},
  \bibnamefont{et~al.}, \bibinfo{journal}{J. Appl. Phys.}
  \textbf{\bibinfo{volume}{104}}, \bibinfo{eid}{043914} (\bibinfo{year}{2008}).

\bibitem[{\citenamefont{Krizakova et~al.}(2019)\citenamefont{Krizakova,
  Pe\~{n}a Garcia, Vogel, de~Souza~Chaves, Pizzini, and
  Thiaville}}]{Krizakova2019}
\bibinfo{author}{\bibfnamefont{V.}~\bibnamefont{Krizakova}},
  \bibinfo{author}{\bibfnamefont{J.}~\bibnamefont{Pe\~{n}a Garcia}},
  \bibinfo{author}{\bibfnamefont{J.}~\bibnamefont{Vogel}},
  \bibinfo{author}{\bibfnamefont{D.}~\bibnamefont{de~Souza~Chaves}},
  \bibinfo{author}{\bibfnamefont{S.}~\bibnamefont{Pizzini}}, \bibnamefont{and}
  \bibinfo{author}{\bibfnamefont{A.}~\bibnamefont{Thiaville}},
  \bibinfo{journal}{Phys. Rev. B} \textbf{\bibinfo{volume}{100}},
  \bibinfo{pages}{214404} (\bibinfo{year}{2019}).

\bibitem[{\citenamefont{Peña~Garcia et~al.}(2021)\citenamefont{Peña~Garcia,
  Fassatoui, Bonfim, Vogel, Thiaville, and Pizzini}}]{PenaGarcia2021}
\bibinfo{author}{\bibfnamefont{J.}~\bibnamefont{Peña~Garcia}},
  \bibinfo{author}{\bibfnamefont{A.}~\bibnamefont{Fassatoui}},
  \bibinfo{author}{\bibfnamefont{M.}~\bibnamefont{Bonfim}},
  \bibinfo{author}{\bibfnamefont{J.}~\bibnamefont{Vogel}},
  \bibinfo{author}{\bibfnamefont{A.}~\bibnamefont{Thiaville}},
  \bibnamefont{and} \bibinfo{author}{\bibfnamefont{S.}~\bibnamefont{Pizzini}},
  \bibinfo{journal}{Phys. Rev. B} \textbf{\bibinfo{volume}{104}},
  \bibinfo{pages}{014405} (\bibinfo{year}{2021}).

\bibitem[{\citenamefont{Katayama et~al.}(1978)\citenamefont{Katayama, Hasegawa,
  Kawanishi, and Tsushima}}]{Katayama1978}
\bibinfo{author}{\bibfnamefont{T.}~\bibnamefont{Katayama}},
  \bibinfo{author}{\bibfnamefont{K.}~\bibnamefont{Hasegawa}},
  \bibinfo{author}{\bibfnamefont{K.}~\bibnamefont{Kawanishi}},
  \bibnamefont{and} \bibinfo{author}{\bibfnamefont{T.}~\bibnamefont{Tsushima}},
  \bibinfo{journal}{J. Appl. Phys.} \textbf{\bibinfo{volume}{49}},
  \bibinfo{pages}{1759} (\bibinfo{year}{1978}).

\bibitem[{\citenamefont{Caretta et~al.}(2018)\citenamefont{Caretta, Mann,
  Buettner, Ueda, Pfau, Guenther, Hessing, Churikoval, Klose, Schneider
  et~al.}}]{Caretta2018}
\bibinfo{author}{\bibfnamefont{L.}~\bibnamefont{Caretta}},
  \bibinfo{author}{\bibfnamefont{M.}~\bibnamefont{Mann}},
  \bibinfo{author}{\bibfnamefont{F.}~\bibnamefont{Buettner}},
  \bibinfo{author}{\bibfnamefont{K.}~\bibnamefont{Ueda}},
  \bibinfo{author}{\bibfnamefont{B.}~\bibnamefont{Pfau}},
  \bibinfo{author}{\bibfnamefont{C.~M.} \bibnamefont{Guenther}},
  \bibinfo{author}{\bibfnamefont{P.}~\bibnamefont{Hessing}},
  \bibinfo{author}{\bibfnamefont{A.}~\bibnamefont{Churikoval}},
  \bibinfo{author}{\bibfnamefont{C.}~\bibnamefont{Klose}},
  \bibinfo{author}{\bibfnamefont{M.}~\bibnamefont{Schneider}},
  \bibnamefont{et~al.}, \bibinfo{journal}{Nat. Nanotechnol.}
  \textbf{\bibinfo{volume}{13}}, \bibinfo{pages}{1154} (\bibinfo{year}{2018}).

\bibitem[{\citenamefont{Vansteenkiste et~al.}(2014)\citenamefont{Vansteenkiste,
  Leliaert, Dvornik, Helsen, Garcia-Sanchez, and
  Van~Waeyenberge}}]{Vansteenkiste2014}
\bibinfo{author}{\bibfnamefont{A.}~\bibnamefont{Vansteenkiste}},
  \bibinfo{author}{\bibfnamefont{J.}~\bibnamefont{Leliaert}},
  \bibinfo{author}{\bibfnamefont{M.}~\bibnamefont{Dvornik}},
  \bibinfo{author}{\bibfnamefont{M.}~\bibnamefont{Helsen}},
  \bibinfo{author}{\bibfnamefont{F.}~\bibnamefont{Garcia-Sanchez}},
  \bibnamefont{and}
  \bibinfo{author}{\bibfnamefont{B.}~\bibnamefont{Van~Waeyenberge}},
  \bibinfo{journal}{AIP Advances} \textbf{\bibinfo{volume}{4}},
  \bibinfo{pages}{107133} (\bibinfo{year}{2014}).

\bibitem[{\citenamefont{Raymenants et~al.}(2021)\citenamefont{Raymenants, Wan,
  Couet, Souriau, Thiam, Tsvetanova, Canvel, Garello, Kar, Heyns
  et~al.}}]{Raymenants2021}
\bibinfo{author}{\bibfnamefont{E.}~\bibnamefont{Raymenants}},
  \bibinfo{author}{\bibfnamefont{D.}~\bibnamefont{Wan}},
  \bibinfo{author}{\bibfnamefont{S.}~\bibnamefont{Couet}},
  \bibinfo{author}{\bibfnamefont{L.}~\bibnamefont{Souriau}},
  \bibinfo{author}{\bibfnamefont{A.}~\bibnamefont{Thiam}},
  \bibinfo{author}{\bibfnamefont{D.}~\bibnamefont{Tsvetanova}},
  \bibinfo{author}{\bibfnamefont{Y.}~\bibnamefont{Canvel}},
  \bibinfo{author}{\bibfnamefont{K.}~\bibnamefont{Garello}},
  \bibinfo{author}{\bibfnamefont{G.~S.} \bibnamefont{Kar}},
  \bibinfo{author}{\bibfnamefont{M.}~\bibnamefont{Heyns}},
  \bibnamefont{et~al.}, \bibinfo{journal}{IEEE Trans. Elec.
  Dev.} \textbf{\bibinfo{volume}{68}}, \bibinfo{pages}{2116}
  (\bibinfo{year}{2021}).
  
  \bibitem[{\citenamefont{Vansteenkiste et~al.}(2014)\citenamefont{Vansteenkiste,
  Leliaert, Dvornik, Helsen, Garcia-Sanchez, and
  Van~Waeyenberge}}]{Vansteenkiste2014}
\bibinfo{author}{\bibfnamefont{A.}~\bibnamefont{Vansteenkiste}},
  \bibinfo{author}{\bibfnamefont{J.}~\bibnamefont{Leliaert}},
  \bibinfo{author}{\bibfnamefont{M.}~\bibnamefont{Dvornik}},
  \bibinfo{author}{\bibfnamefont{M.}~\bibnamefont{Helsen}},
  \bibinfo{author}{\bibfnamefont{F.}~\bibnamefont{Garcia-Sanchez}},
  \bibnamefont{and}
  \bibinfo{author}{\bibfnamefont{B.}~\bibnamefont{Van~Waeyenberge}},
  \bibinfo{journal}{AIP Advances} \textbf{\bibinfo{volume}{4}},
  \bibinfo{pages}{107133} (\bibinfo{year}{2014}).

\bibitem[{\citenamefont{Haltz et~al.}(2021)\citenamefont{Haltz, Krishnia,
  Berges, Mougin, and Sampaio}}]{Haltz2021}
\bibinfo{author}{\bibfnamefont{E.}~\bibnamefont{Haltz}},
  \bibinfo{author}{\bibfnamefont{S.}~\bibnamefont{Krishnia}},
  \bibinfo{author}{\bibfnamefont{L.}~\bibnamefont{Berges}},
  \bibinfo{author}{\bibfnamefont{A.}~\bibnamefont{Mougin}}, \bibnamefont{and}
  \bibinfo{author}{\bibfnamefont{J.}~\bibnamefont{Sampaio}},
  \bibinfo{journal}{Phys. Rev. B} \textbf{\bibinfo{volume}{103}}
  (\bibinfo{year}{2021}).

\bibitem[{\citenamefont{Krizakova et~al.}(2019)\citenamefont{Krizakova,
  Pe\~{n}a Garcia, Vogel, de~Souza~Chaves, Pizzini, and
  Thiaville}}]{Krizakova2019}
\bibinfo{author}{\bibfnamefont{V.}~\bibnamefont{Krizakova}},
  \bibinfo{author}{\bibfnamefont{J.}~\bibnamefont{Pe\~{n}a Garcia}},
  \bibinfo{author}{\bibfnamefont{J.}~\bibnamefont{Vogel}},
  \bibinfo{author}{\bibfnamefont{D.}~\bibnamefont{de~Souza~Chaves}},
  \bibinfo{author}{\bibfnamefont{S.}~\bibnamefont{Pizzini}}, \bibnamefont{and}
  \bibinfo{author}{\bibfnamefont{A.}~\bibnamefont{Thiaville}},
  \bibinfo{journal}{Phys. Rev. B} \textbf{\bibinfo{volume}{100}},
  \bibinfo{pages}{214404} (\bibinfo{year}{2019}).

\bibitem[{\citenamefont{Peña~Garcia et~al.}(2021)\citenamefont{Peña~Garcia,
  Fassatoui, Bonfim, Vogel, Thiaville, and Pizzini}}]{PenaGarcia2021}
\bibinfo{author}{\bibfnamefont{J.}~\bibnamefont{Peña~Garcia}},
  \bibinfo{author}{\bibfnamefont{A.}~\bibnamefont{Fassatoui}},
  \bibinfo{author}{\bibfnamefont{M.}~\bibnamefont{Bonfim}},
  \bibinfo{author}{\bibfnamefont{J.}~\bibnamefont{Vogel}},
  \bibinfo{author}{\bibfnamefont{A.}~\bibnamefont{Thiaville}},
  \bibnamefont{and} \bibinfo{author}{\bibfnamefont{S.}~\bibnamefont{Pizzini}},
  \bibinfo{journal}{Phys. Rev. B} \textbf{\bibinfo{volume}{104}},
  \bibinfo{pages}{014405} (\bibinfo{year}{2021}).

\bibitem[{\citenamefont{Leliaert et~al.}(2014)\citenamefont{Leliaert, Van~de
  Wiele, Vansteenkiste, Laurson, Durin, Dupr{\'e}, and
  Van~Waeyenberge}}]{leliaert2014current}
\bibinfo{author}{\bibfnamefont{J.}~\bibnamefont{Leliaert}},
  \bibinfo{author}{\bibfnamefont{B.}~\bibnamefont{Van~de Wiele}},
  \bibinfo{author}{\bibfnamefont{A.}~\bibnamefont{Vansteenkiste}},
  \bibinfo{author}{\bibfnamefont{L.}~\bibnamefont{Laurson}},
  \bibinfo{author}{\bibfnamefont{G.}~\bibnamefont{Durin}},
  \bibinfo{author}{\bibfnamefont{L.}~\bibnamefont{Dupr{\'e}}},
  \bibnamefont{and}
  \bibinfo{author}{\bibfnamefont{B.}~\bibnamefont{Van~Waeyenberge}},
  \bibinfo{journal}{J. Appl. Phys.} \textbf{\bibinfo{volume}{115}},
  \bibinfo{pages}{233903} (\bibinfo{year}{2014}).


\end{thebibliography}
\end{document}